\PassOptionsToPackage{dvipsnames}{xcolor}
  \documentclass[year=26,pdfa]{fmcad}
  
  \usepackage{cite}
  \usepackage{amsmath,amssymb,amsfonts,amsthm}
  \usepackage{graphicx}
  \usepackage{textcomp}
  \usepackage[dvipsnames]{xcolor}

  \usepackage{bussproofs}  
  \usepackage{mathpartir}
  \usepackage{orcidlink}
  \usepackage{booktabs}
  \usepackage{siunitx}
  \usepackage{float}       
  \usepackage{placeins}    
  \usepackage{xspace}      
  \usepackage{multirow}    
  \usepackage{enumerate}   
  \usepackage[square,comma,numbers,sort&compress]{natbib}      
  \usepackage{diagbox}
  \usepackage{booktabs}
  \usepackage{pifont}
  \usepackage{tikz}
  \usetikzlibrary{matrix}

  \usepackage{arydshln}

  \newcommand*\dotdiam{15pt}      
  \newcommand*\dotgap {4pt}      

  \tikzset{
    cell/.style   = {draw,circle,fill=gray!60,
                    minimum size=\dotdiam, inner sep=0pt, outer sep=0pt,
                    text height=1.5ex, text depth=0.5ex},
    redcell/.style= {cell,fill=red!60},
    redpluscell/.style= {cell,fill=red!60,execute at begin node={\raisebox{-1pt}{\Large +}}},
    redminuscell/.style= {cell,fill=red!60,execute at begin node={\raisebox{-1pt}{\Large -}}},
    greencell/.style  = {cell,fill=green!60},
    greenpluscell/.style  = {cell,fill=green!60,execute at begin node={\raisebox{-1pt}{\Large +}}},
    greenminuscell/.style  = {cell,fill=green!60,execute at begin node={\raisebox{-1pt}{\Large -}}},
    graycell/.style = {cell,fill=gray!40},
    graypluscell/.style = {cell,fill=gray!40,execute at begin node={\raisebox{-1pt}{\Large +}}},
    grayminuscell/.style = {cell,fill=gray!40,execute at begin node={\raisebox{-1pt}{\Large -}}},
    bluecell/.style = {cell,fill=blue!20},
    bluepluscell/.style = {cell,fill=blue!20,execute at begin node={\raisebox{-1pt}{\Large +}}},
    blueminuscell/.style = {cell,fill=blue!20,execute at begin node={\raisebox{-1pt}{\Large -}}},
    yellowcell/.style = {cell,fill=yellow!60},
    yellowpluscell/.style = {cell,fill=yellow!60,execute at begin node={\raisebox{-1pt}{\Large +}}},
    yellowminuscell/.style = {cell,fill=yellow!60,execute at begin node={\raisebox{-1pt}{\Large -}}},
    purplecell/.style = {cell,fill=purple!60},
    purplepluscell/.style = {cell,fill=purple!60,execute at begin node={\raisebox{-1pt}{\Large +}}},
    purpleminuscell/.style = {cell,fill=purple!60,execute at begin node={\raisebox{-1pt}{\Large -}}},
    orangecell/.style = {cell,fill=orange!60},
    orangepluscell/.style = {cell,fill=orange!60,execute at begin node={\raisebox{-1pt}{\Large +}}},
    orangeminuscell/.style = {cell,fill=orange!60,execute at begin node={\raisebox{-1pt}{\Large -}}},
    gridmatrix/.style = {
        matrix of nodes,
        nodes   = {cell},
        column sep=\dotgap, row sep=\dotgap,
        nodes in empty cells,                          
    },
  }
  \usepackage{subcaption}
  
  \usepackage{outlines}
  \usepackage{listings}
  \usepackage{soul}
  \usepackage{microtype}
  
  \def\BibTeX{{\rm B\kern-.05em{\sc i\kern-.025em b}\kern-.08em
      T\kern-.1667em\lower.7ex\hbox{E}\kern-.125emX}}
  
  \usepackage[linesnumbered,ruled,vlined]{algorithm2e}
  \usepackage{hyperref}
  \hypersetup{
    colorlinks=true,     
    urlcolor=blue,       
    linkcolor=blue,     
    citecolor=blue,     
    plainpages=false,
    breaklinks=true,
    bookmarksnumbered=true,
    bookmarksopen=true,
    pdftitle={},
    pdfauthor={},
    pdfkeywords={},
    pdfpagemode=UseOutlines  
  }

  \usepackage[capitalise]{cleveref}
  \usepackage{tabularx}
  \usepackage{enumitem}

\begin{document}
  \title{Parallel SMT Solving via Dynamic Partitioning, Core-Guided Pruning, and Backbone Detection}

\author{
Ilana Shapiro\textsuperscript{1} \orcidlink{0000-0003-0848-0486}, 
Sorin Lerner\textsuperscript{2,1} \orcidlink{0000-0003-3957-0628},
Nikolaj Bj{\o}rner\textsuperscript{3} \orcidlink{0000-0002-1695-2810}
\\
\textsuperscript{1}\emph{UC San Diego, La Jolla, CA, USA} \\
\textsuperscript{2}\emph{Cornell University, Ithaca, NY, USA} \\
\textsuperscript{3}\emph{Microsoft Research, Redmond, WA, USA} \\
\texttt{\small ilshapiro@ucsd.edu, sorin.lerner@cornell.edu, nbjorner@microsoft.com}
}
  
  \maketitle
  
  \begin{abstract}
   Exploiting parallelism in modern CPU architectures remains a longstanding challenge in optimizing SMT solvers. We introduce a novel framework for parallel SMT solving that uses \emph{feedback from active search} to steer solving. We dynamically build a \textit{binary partition tree} of the search space by sampling from workers' VSIDS statistics during solving. We introduce a novel search-space pruning mechanism that harnesses the full power of \textit{core-based CDCL-style pruning} to continuously shrink the partition tree. We further optimize our architecture by incorporating \textit{online backbone detection} into worker threads, as well as a \textit{terminate-on-demand} mechanism to eagerly eliminate work on pruned subproblems. The resulting algorithm is highly generalizable and scales effectively with available resources. We implement our approach in the \textsc{z3} SMT solver and demonstrate that it outperforms both sequential \textsc{z3} and existing state-of-the-art parallel frameworks on challenging benchmarks from six logics in the SMT-COMP 2025 Parallel Track.
  \end{abstract}
  \section{Introduction}~\label{sec:intro}
Scaling distributed SMT to match rapid advances in multi-core CPUs, high-performance computing, and cloud platforms is a longstanding challenge. Ideally, solver performance should scale with available physical resources. However, most state-of-the-art SMT solvers, such as \textsc{z3} \cite{z3}, \textsc{cvc5} \cite{cvc5}, \textsc{OpenSMT} \cite{opensmt2}, and \textsc{Yices2} \cite{yices2.2}, remain largely single-threaded. Consequently, most advances in solver performance have focused on improving the techniques and heuristics of sequential SMT solvers, leaving substantial computational resources underexploited. This limitation constrains the effectiveness of SMT solvers in key domains such as verification \cite{verification}, model checking \cite{model_checking}, security \cite{security}, and program synthesis \cite{synthesis}. 

Existing parallel approaches broadly fall into two categories: portfolio and partitioning. In the portfolio method, multiple solvers are run in parallel under different configurations and/or random seeds and may exchange learned clauses during solving. The runtime of SMT solvers is highly variable and sensitive to seemingly trivial perturbations of the solver \cite{smts2026, mariposa_smt_runtime}. Portfolio solving has led to substantial improvements over sequential approaches by exploiting this runtime instability through parallel sampling of the solver configuration space \cite{smt-d, z3_oldparallel, parallel-sat-handbook}. However, portfolio solving is hindered by limited scalability since it is bounded by the performance of the best sequential solver \cite{cvc5_partition, smts2026}. The partitioning approach addresses scalability via a divide-and-conquer strategy: the input formula is decomposed into independent subproblems whose disjunction is a tautology to the original formula. The goal is to shrink the search space of each subproblem such that solving them in parallel is faster than sequential. An optimal search space partitioning has the potential to vastly outperform portfolio solving. However, the partitioning strategy faces several inherent challenges: it is difficult to partition the search space into subproblems that are easier than the original, potentially wasting resources to solve infeasible sub-problems when the original formula is satisfiable.

A popular approach to partitioning is \textit{cube-and-conquer}, in which $n$ Boolean atoms are chosen from the input formula and assigned to both polarities. This creates $2^n$ independent cubes (partial assignments) for parallel solving. Selecting good splitting atoms is difficult, and badly selected atoms can produce a harder problem than the original, particularly when the partitions are static \cite{parallel-sat-handbook, static_cubing_proof_prefixes}. Recent techniques have turned to \textit{dynamic} partitioning \cite{cvc5_partition, ariparti, smts2026}, where split atoms are selected during search. A similarly challenging problem is cube scheduling across threads, as the order of subproblem exploration can substantially impact overall performance. The \textit{partition tree} has shown success as a data structure for dynamic partitioning. Recent architectures have combined it with clause sharing and basic search-space pruning \cite{smts_tree_algs_2015, smts2026}. However, none of these approaches fully leverage feedback from active search.

In this paper, we move toward a more robust dynamic partitioning framework by systematically \emph{exploiting feedback from active search} to guide both partitioning and pruning decisions. We introduce a binary partition tree whose splitting atoms are selected on-the-fly from worker threads’ VSIDS statistics. By leveraging the conflict-driven reasoning of CDCL, we enable non-chronological backjumping across the partition tree, using per-thread unsatisfiable cores to actively prune the search space. Our partition tree also supports cross-thread information sharing and on-demand termination of workers on shared subproblems. We further augment our framework with core minimization and online backbone detection for additional search-space pruning. We implement our approach in \textsc{z3}, and show that our approach outperforms both sequential \textsc{z3} and state-of-the-art parallel frameworks on challenging benchmarks from six logics in the SMT-COMP 2025 Parallel Track.

In summary, we make the following contributions: 

\begin{enumerate} 
    \item We introduce a generalizable framework for parallel SMT solving that \textit{dynamically partitions} the input formula with Boolean atoms sampled on-the-fly from VSIDS statistics.
    \item We introduce a \textit{binary partition tree} for cube distribution with a novel pruning mechanism that harnesses the full power of \textit{core-based CDCL-style pruning} to dynamically shrink the problem space.
    \item We further optimize our architecture by incorporating \textit{core minimization} and \textit{online backbone detection} for more powerful search-space pruning.
    \item We implement our tool in the SMT solver \textsc{z3} and show that our approach outperforms sequential \textsc{z3} and existing parallel frameworks on six logics from the SMT-COMP 2025 Parallel Track. Our implementation, data, and results are available at
    \url{https://zenodo.org/records/21755875}
\end{enumerate}

  \section{Preliminaries}~\label{sec:preliminaries}

\subsection{Definitions and Notations}
\newcommand{\Bool}{\mathsf{BOOL}}

Satisfiability Modulo Theories (SMT) determine the satisfiability of a first-order logic formula with respect to a background theory. A \emph{theory} is a pair $T = (\Sigma, I)$ where $\Sigma$ is a signature and $I$ is a class of $\Sigma$-interpretations. We assume a fixed background theory $T$ with signature $\Sigma$ that includes the Boolean sort $\Bool$, and assume that all terms are $\Sigma$-terms, entailment ($\models$) is entailment modulo $T$, equivalence is equivalence modulo $T$, and interpretations are $T$-interpretations. A \emph{model} of a formula $\varphi$ is a $T$-interpretation $\mathcal{M}$ such that $\mathcal{M} \models \varphi$. An \emph{implicant} of $\varphi$ is a conjunction of literals $\nu$ such that $\nu \models \varphi$, i.e. every model of $\nu$ is also a model of $\varphi$. An \emph{atom} is a term of sort $\Bool$ that contains no subterms of this sort. A \emph{literal} is an atom or its negation. A \emph{clause} is a disjunction of literals, and a \emph{cube} is a conjunction of literals. A propositional formula in conjunctive normal form (CNF) is a conjunction of clauses. A formula $\varphi$ is a term of sort $\Bool$ and is satisfiable if it is satisfied by some interpretation in I, and unsatisfiable otherwise. A formula whose negation is unsatisfiable is \emph{valid} \cite{sat_handbook}.

Most SMT solvers follow the CDCL($T$) framework~\cite{dpll_t}, which combines a CDCL SAT solver with one or more theory solvers. The input formula is first preprocessed into an equisatisfiable CNF formula $\phi$ preserving its $T$-semantics. The SAT solver then incrementally constructs an assignment $\alpha$ through alternating \emph{propagation} (unit clauses force assignments to literals) and \emph{decision} (solver chooses values for unassigned literals) phases. In the decision phase, the Variable State Independent Decaying Sum (VSIDS) heuristic typically guides branching decisions by prioritizing variables involved in recent conflicts. These assignments and their dependencies are tracked in an \emph{implication graph} \cite{implication_graph}. The SAT solver performs conflict analysis to learn a clause via resolution and adds it to its clause database to prune the search space. 
During conflict analysis, the solver may extract an \emph{unsatisfiable core}: a subset of assigned literals inconsistent with respect to $T$.
The process repeats until a $T$-consistent satisfying assignment is found (SAT) or an unrecoverable conflict is derived (UNSAT)~\cite{sat_handbook}.

\subsection{Portfolio Solving} \label{subsec:portfolio_solving}

SMT solvers are extremely sensitive to small perturbations to the input formula or configuration heuristics \cite{parallel-sat-handbook}. The \emph{portfolio} strategy leverages this intrinsic instability by running multiple solvers in parallel with slightly different configurations (e.g. parameter values, random seeds) or with permuted, but logically equivalent, inputs. Solvers exchange information learned during search. The portfolio approach has made substantial gains over sequential solving \cite{smt-d, z3_oldparallel, HordeSat, smts_2016, smts2018}. However, since the fastest solver determines the performance of the portfolio, increasing parallelism leads to diminishing returns. 

\subsection{Search-Space Partitioning} \label{subsec:search_space_partitioning}

The \emph{partitioning} approach divides $\varphi$ into $n$ independent subproblems
$\varphi_1, \dots, \varphi_n$ such that $\varphi \equiv \bigvee_i \varphi_i$. Thus, if any $\varphi_i$ is SAT, then $\varphi$ is SAT, and if all $\varphi_i$ are UNSAT, then $\varphi$ is UNSAT. There are two main partitioning strategies: \emph{cube-and-conquer} and \emph{scattering}. In cube-and-conquer, a set of $n$ splitting atoms $A = \{a_1, \dots, a_n\}$ is selected. Traditionally, heuristics such as lookahead (which ranks candidate literals by how much their tentative assignments simplify the current subproblem)
and VSIDS (which prioritizes variables in recent conflict activity) are used for split atom selection \cite{lookahead_partitioning, parallel-sat-handbook}. A \emph{cube} is a conjunction of literals over these atoms, i.e. a formula of the form $C = \ell_1 \land \cdots \land \ell_n$, where each $\ell_i \in \{ a_j, \neg a_j \}$ for some $a_j \in A$. The $2^n$ resulting cubes define $2^n$ independent subproblems $\varphi \land C$ for parallel solving. 

In contrast, scattering produces $n$ partitioning formulae:
\begin{align*}
p_1 &= C_1, \\
p_2 &= \neg C_1 \land C_2, \\
p_3 &= \neg C_1 \land \neg C_2 \land C_3, \\
&\ \ \vdots \\
p_n &= \neg C_1 \land \cdots \land \neg C_{n-1} \land C_n,
\end{align*}
where each $C_i$ is a cube (not necessarily over the same set of atoms). Each partition induces a subproblem of the form $\varphi \land p_i$. By construction, all $p_i$ are disjoint and their disjunction $\bigvee_{i=1}^n p_i$ covers the search space. Thus, $\varphi \equiv \bigvee_{i=1}^n (\varphi \land p_i)$. 

A popular data structure for search-space partitioning is the \emph{partition tree} \cite{smts_tree_algs_2015}, where the root node represents the input formula $\varphi$. Given a parent node associated with formula $P$, its $i$-th child represents a formula $P \land C_i$ produced by a \emph{partitioning function} \cite{partitioning_smts_2020} such that for all $i$, the disjunction $\bigvee_{i=1}^n C_i$ holds, and for all $j \neq i$, $\neg (C_i \land C_j)$. Search-space partitioning can vastly outperform portfolio if good partitions are chosen. This is a challenging task, and poorly chosen partitions can result in harder problems than the original \cite{lookahead_partitioning}.
  \section{Related Work}~\label{sec:related_work}

\textbf{Portfolio SMT Solving.}
Portfolio SMT solving with clause sharing was first introduced in the \textsc{z3} SMT solver \cite{z3_oldparallel}, which simply randomized 4 instances of \textsc{z3} and shared globally valid lemmas with at most 8 literals between them. 
Since \textsc{z3}'s initial attempt at portfolio parallelism, several portfolio architectures have been pursued \cite{par4, yices2.2}. The most successful frameworks were developed in \textsc{smts} \cite{smts_tree_algs_2015, smts_2016, smts2018} and \textsc{cvc5} \cite{smt-d}. \textsc{smts} adopts a unique client-server architecture that orchestrates an underlying solver. Globally valid lemmas are stored in an external database, scored by frequency, and randomly sampled. Lemmas with more than 3 literals are discarded \cite{smts2026}. \textsc{cvc5} provides further optimizations such as delayed clause sharing and guided randomization \cite{smt-d}. Although portfolio solving has led to significant speedups (particularly in \cite{smt-d} and \cite{smts_2016}), it still suffers from innate lack of scalability.

\textbf{SMT Search-Space Partitioning.} The state-of-the-art SMT search-space partitioning framework is \textsc{smts}. It is the first to combine partitioning with clause sharing and per-partition portfolios, and the first to revisit already attempted subproblems with a selection strategy rooted in the solver runtime random distribution \cite{smts_tree_algs_2015, smts_2016, smts2018, smts2026}. \textsc{smts}'s underlying solver is \textsc{OpenSMT2}, which uses the scattering strategy to construct partitions from CDCL search branch decision literals \cite{smts_scattering, opensmt2}. 
Central to \textsc{SMTS}'s architecture is the \emph{partitioning tree} (\cite{smts_tree_algs_2015}), which is built dynamically during search \cite{smts2026}. 

In contrast to \textsc{smts}, \textsc{cvc5-Cloud} does not propose a new framework; rather, it introduces techniques for combining multiple atom sources and partitioning with these atoms \cite{cvc5_partition}. \textsc{AriParti} maintains a dynamic partition tree similar to \textsc{smts}, but performs variable-level partitioning tailored to arithmetic theories, simplifying subproblems using Boolean and Interval Constraint Propagation \cite{ariparti}.
Unlike \textsc{smts}, these works do not support clause sharing or revisit already attempted instances. Finally, other partitioning frameworks either rely on \emph{static} partitioning trees \cite{static_cubing_proof_prefixes, smts_2016}, which are less effective than a carefully constructed dynamic tree, or are tailored to specific logics (\textsc{Bitwuzla} \cite{Bitwuzla} and \textsc{PBoolector} \cite{PBoolector}, which combine cube-and-conquer with bitblasting for \textsf{QF-BV}) and thus have limited generality. While \textsc{smts} and \textsc{AriParti} support basic search-space pruning, more sophisticated strategies remain unexplored. Existing frameworks also do not fully incorporate feedback from search. Our framework addresses both of these gaps.

\textbf{Parallel SAT Solving.} The state-of-the-art parallel SAT framework \textsc{Mallob} uses \emph{malleable scheduling} to allocate progressively more resources to challenging tasks while preserving state across incremental SAT queries. The authors show that slight differences in learned-clause import timing cause initially identical solver threads to diverge, enabling effective search pruning without explicit portfolio diversification or search-space partitioning \cite{mallob,mallobsat}. \textsc{Mallob} has recently been applied to parallel bit-vector SMT solving in \textsc{Bitwuzllob} \cite{bitwuzllob}, which replaces \textsc{Bitwuzla}'s \cite{Bitwuzla} sequential CDCL backend with \textsc{Mallob}. This integration exploits \textsc{Bitwuzla}'s offline SAT architecture, where Boolean reasoning is delegated to complete incremental SAT calls. However, integrating \textsc{Mallob} into a general CDCL($T$)-based SMT solver is not straightforward, as CDCL($T$) requires fine-grained interaction between the Boolean and theory solvers during search \cite{bitwuzllob}. \textsc{Mallob}'s IPASIR-based interface submits clauses and assumptions for a complete SAT solving call but provides no callbacks through which a theory solver can inspect partial assignments or return propagations and conflicts during search \cite{mallob_ipasir}. IPASIR-UP~\cite{FazekasNPKSB23} provides callbacks suitable for implementing this interaction but is not currently supported by \textsc{Mallob}.
  \section{Parallel Architecture}~\label{sec:implementation}

\begin{figure}[!t] 
    \centering
    \includegraphics[width=0.3\textwidth]{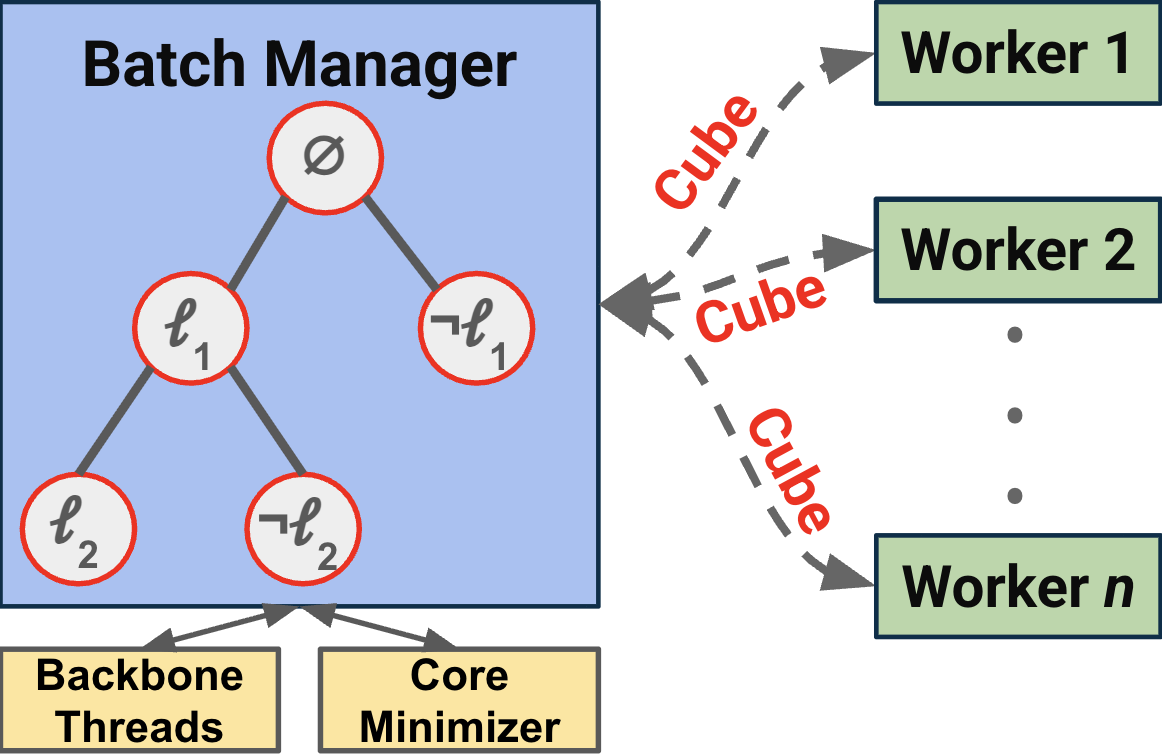}
    \caption{Overview of Parallel Architecture}
    \label{fig:architecture}
\end{figure}

In this section, we detail our novel parallel architecture and the strategies it employs to dynamically leverage information from search. Our motivating design principle is \emph{leveraging feedback from active search}, with our final architecture refined through empirical evaluation (Figure \ref{fig:architecture}).  
Our approach is based on cube-and-conquer, with cubes are created on-the-fly by sampling split atoms from solver threads' VSIDS statistics\footnote{VSIDS was selected over alternative heuristics, including lookahead, based on experimental signals}. On the main thread, the \emph{batch manager} coordinates thread synchronization. It maintains a dynamically evolving \emph{binary partition tree} for cube storage, distributes cubes to $n$ worker threads, and facilitates information exchange between workers. Each worker $w$ runs a sequential \textsc{z3} instance with a progressively increasing conflict budget and repeatedly requests cubes from the batch manager. If the cube is SAT, the entire problem is SAT. If $w$ exhausts its conflict budget without solving the cube, it returns control to the batch manager, which may perform subcubing, and selects a new cube for $w$ (Section~\ref{subsec:searchtree}). If the cube is UNSAT, the batch manager prunes the partition tree using the solver's UNSAT core (Section~\ref{subsec:pruning}) and similarly assigns $w$ a new cube. To mitigate redundant computation when multiple workers are assigned to the same cube, we employ a \emph{terminate-on-demand} policy (Section \ref{subsec:terminate_on_demand}) that aborts workers operating on now-stale cubes. Throughout execution, $w$ accumulates learned lemmas that are reused across cubes. Finally, outside the partition tree, the batch manager coordinates two \emph{backbone threads}, which prove backbone literals detected during search (Section~\ref{subsec:failed_literals}) to shrink the search space of the worker threads, as well as a dedicated \emph{core minimization thread} (Section~\ref{subsec:core_min}), which asynchronously reduces workers' UNSAT cores for more powerful pruning.


\subsection{Binary Partition Tree}~\label{subsec:searchtree}

At the center of our algorithm is the \emph{partition tree}. Unlike \textsc{smts} and \textsc{AriParti}, which support $n$-ary partition trees, we adopt a binary tree to facilitate our core-guided pruning algorithm (Section~\ref{subsec:pruning}). Each node $n$ in our partition tree contains a \emph{split atom} of positive or negative polarity (i.e. $\ell$ or $\neg \ell$), and the children of $n$ are the polarities of the next split atom (i.e. $\ell_1$ and $\neg \ell_1$) (Figure \ref{fig:architecture}). Each node $n$ encodes a cube $C$ via the path of literals from $n$ to root. When a worker is assigned node $n$, it solves the formula $\varphi \land C$, where $\varphi$ is the input formula, under a designated conflict budget. The partition tree is initialized with the empty cube $\emptyset$ at the root (Figure \ref{fig:architecture}). Thus, all workers begin with $\varphi$ itself. 

Nodes exist in one of three states: \texttt{open} (unsolved and has no assigned workers), \texttt{active} (has at least 1 assigned worker), or \texttt{closed} (was determined UNSAT). After solving their cube or hitting their conflict budget, workers report one of 3 statuses to the batch manager: SAT, UNSAT, and UNDEF. A SAT result terminates the search ($\varphi$ is SAT). An UNSAT result marks the node as \texttt{closed}: the batch manager prunes the partition tree (Section \ref{subsec:pruning}), and shares the worker's UNSAT core with the other workers. If the entire tree is closed, search is terminated with an UNSAT result. 

When worker $w$ reports UNDEF on node $n$, the batch manager decides (1) if $n$ should be split, (governed by the \emph{tree expansion policy}), and (2) $w$'s next node assignment (given by the \emph{node selection policy}). Globally valid unit clauses learned by $w$ on $n$ are shared with the other workers. We do not share larger lemmas at this time; this remains a direction for future work. $w_i$'s conflict budget (initially 1000) increases dynamically by a factor of 1.5 after each UNDEF result. Finally, nodes accumulate \emph{effort} from workers who report UNDEF. A worker $w_i$'s effort is its current conflict budget; thus, effort is scaled dynamically. Multiple workers with distinct solver states (VSIDS scores, learned lemmas, etc) and random seeds may be assigned to the same (\texttt{active}) node $n$ in \emph{portfolio} if no \texttt{open} nodes remain. In this case, $n$ only records the maximum effort among its workers to avoid disproportionately inflating the accumulated effort of nodes with higher parallelism. 

Our tree expansion policy (Algorithm \ref{alg:tree_expansion}) draws from \cite{smts2026}. When $w$ reports UNDEF for node $n$, the batch manager counts the \texttt{active} and unsolved (\texttt{open} or \texttt{active}) nodes in the tree. We expand $n$ only if the number of unsolved nodes is less than twice the number of \texttt{active} nodes, all \texttt{open} nodes have been visited at least once, and the depth of $n$ is equal to the depth of the shallowest timed-out leaf. Expansion is then randomly throttled (50\% chance). As noted in \cite{smts2026}, such heuristics ensure the tree remains largely balanced. If expansion proceeds, we split $n$ on the highest scoring atom from $w_i$'s VSIDS statistics. Our node selection policy (Algorithm \ref{alg:node_selection}) is also inspired by \cite{smts2026}. We first search for an \texttt{open} node (target status = \texttt{open}). If none remain, we select from \texttt{active} nodes (target status = \texttt{active}). Among nodes of the chosen status, we select the one with lowest accumulated effort, with greater depth as the tiebreaker. After this, order is random. Although inspired by \cite{smts2026}, Algorithms \ref{alg:tree_expansion} and \ref{alg:node_selection} were empirically adapted to our framework, with some heuristics (e.g. large-tree expansion throttling) omitted due to negative experimental signals. 

\newcommand{\bestEffort}{\mathit{bestEffort}}
\newcommand{\bestDepth}{\mathit{bestDepth}}
\newcommand{\bestNode}{\mathit{bestNode}}
\newcommand{\effort}{\mathit{effort}}
\newcommand{\update}{\mathit{update}}
\newcommand{\target}{\mathit{target}}

\begin{algorithm}[t]
\caption{Tree Expansion (based on \cite{smts2026})}
\label{alg:tree_expansion}

\KwIn{Node $n$; split literal $\ell$; effort $\effort$}

$n.\mathrm{update\_round\_max\_effort}(\effort)$\;

\If{$\neg n.\mathrm{is\_leaf}()$}{
    \Return\;
}

$numActive \gets \mathrm{count\_active\_nodes}(root)$\;
$numUnsolved \gets \mathrm{count\_unsolved\_nodes}(root)$\;

\If{$numUnsolved \ge numActive \cdot 2$}{
    \Return\;
}

\If{$\mathrm{has\_unvisited\_open\_node}(root)$}{
    \Return\;
}
\If{$\mathrm{rand}() \geq 0.5$}{
    \Return\tcp*[r]{50\% rejection}
}

$s \gets \mathrm{shallowest\_timedout\_leaf\_depth}(root)$\;

\If{$\mathrm{depth}(n) = s$}{
    $n.\mathrm{split}(\ell, \neg \ell)$\;
}

\end{algorithm}

\begin{algorithm}[t]
\caption{Node Selection (based on \cite{smts2026})}
\label{alg:node_selection}
\SetKwFunction{DFS}{DFS}
\SetKwProg{Fn}{Function}{}{}

\KwIn{Root node $root$; status $target \in$ \{open, active\}}
\KwOut{Best node $\bestNode$}

$\bestNode \gets \bot; \bestEffort \gets \infty; \bestDepth \gets -1$\;

\DFS{$root$}

\Return $\bestNode$\;

\vspace{0.3em}

\Fn{\DFS{$cur$}}{
    \If{$cur = \bot \vee \mathrm{status}(cur) = \mathrm{closed}$}{
        \Return\;
    }

    \If{$\mathrm{status}(cur) = \target$}{
        $e \gets \mathrm{effort}(cur); d \gets \mathrm{depth}(cur)$\; 
        $update \gets$ false\;

        \If{$\bestNode = \bot$}{
            $\update \gets$ true\;
        }
        \ElseIf{$e \neq \bestEffort$}{
            $\update \gets (e < \bestEffort)$\;
        }
        \Else{
            $\update \gets (d > \bestDepth)$\;
        }
        
        \If{update}{
            $\bestNode \gets cur$\;
            $\bestEffort \gets e$\;
            $\bestDepth \gets d$\;
        }
    }

    \DFS{$\mathrm{left}(cur)$}\;
    \DFS{$\mathrm{right}(cur)$}\;
}
\end{algorithm}


\subsection{Core-Guided Backjumping}~\label{subsec:pruning}

In contrast to prior work \cite{smts2026, ariparti}, when worker $w$ reports UNSAT on node $n$, we do not simply close $n$ and its subtree. Instead, we introduce a novel search-space pruning algorithm that leverages UNSAT cores to perform CDCL-style non-chronological backjumping over the partition tree (Algorithm \ref{alg:backtracking}). If $n$'s UNSAT core is empty, this proves global UNSAT (cubes are passed as assumptions in \textsc{z3}, so an empty core means none were needed). Otherwise, 
all conflict literals lie on the path from the root to $n$ (i.e. are a subset of $w$'s cube).
If $n$'s literal is not in the core, it is irrelevant to the UNSAT result. Thus, we traverse upward from $n$ to the nearest ancestor $n_1$ whose decision literal does appear in the core.\footnote{It is certainly possible that $n_1 = n$.} It is possible that $n_1$ was already closed by another thread with core $C$. In this case, we perform a \emph{core strengthening} check. Let $C'$ be the current core. If $|C'| < |C|$, we replace $C$ with $C'$.

We then attempt to propagate this core upward via \emph{sibling resolution}. Let $\ell$ and $\neg \ell$ be the complementary decision literals of $n_1$ and its sibling, respectively. If both siblings are closed with cores $C_{\ell}$ and $C_{\neg \ell}$, we compute the \emph{sibling resolvent}:\footnote{Both cores are nonempty by construction. An empty core would signal global UNSAT and thus is never attached to a node in an unsolved tree.}
\[
r = (C_\ell \cup C_{\neg \ell}) \setminus \{\ell, \neg \ell\}
\]
If $r = \emptyset$, we conclude global UNSAT. Otherwise, we bubble up to the highest ancestor $a$ such that all literals in $r$ are contained in the path from $a$ to root. We then close $a$ and its subtree, and attach $r$ to $a$ and the newly closed nodes in its subtree. This process is applied recursively: if the sibling of $a$ has already been closed by another thread, we compute a new resolvent and continue propagating upward. The result of this process is non-chronological propagation of cores across the partition tree, enabling integration of independently derived core from different workers and lifting CDCL-style clause resolution to search-space partitioning. We apply this process not only to $n$, but also to other nodes in the partition tree whose cube contains all literals in the UNSAT core.

\begin{algorithm}[t]
\caption{Core-Guided Backjumping}
\label{alg:backtracking}

\KwIn{Node $n$; UNSAT core $C$}

\If{$C = \emptyset$}{
    $\mathrm{close}(root, C)$ \tcp*[r]{global UNSAT}
    \Return\;
}

\While{$n \neq \bot$}{
    \If{$n.\mathrm{literal} \in C$}{
        $\mathrm{CloseWithCore}(n, C)$\;
        \Return\;
    }
    $n \gets n.\mathrm{parent}$\;
}

\vspace{0.3em}

\SetKwFunction{CWC}{CloseWithCore}
\SetKwProg{Fn}{Function}{}{}

\Fn{\CWC{$n, C$}}{
    \If{$n.\mathrm{closed}$}{
        \Return\;
    }

    $p \gets n.\mathrm{parent}$\;
    $\mathrm{close}(n, C)$\;
    $l \gets p.\mathrm{left}$;\quad $r \gets p.\mathrm{right}$\;

    \If{$l.\mathrm{closed}$ $\land$ $r.\mathrm{closed}$}{
        $R \gets (l.\mathrm{core} \cup r.\mathrm{core}) \setminus \{l.\mathrm{lit}, r.\mathrm{lit}\}$\;

        \If{$R = \emptyset$}{
            $\mathrm{close}(root, R)$ \tcp*[r]{global UNSAT}
            \Return\;
        }

        $a \gets \mathrm{highest\_attach}(p, R)$\;
        $\mathrm{close}(a, R)$\;

        $\mathrm{PropagateUpward}(a)$\;
    }
}

\vspace{0.3em}

\SetKwFunction{PU}{PropagateUpward}

\Fn{\PU{$cur$}}{
    \While{$cur.\mathrm{parent} \neq \bot$}{
        $p \gets cur.\mathrm{parent}$; \quad $l \gets p.\mathrm{left}$;\quad $r \gets p.\mathrm{right}$\;

        \If{$\neg (l.\mathrm{closed} \land r.\mathrm{closed})$}{
            \Return\;
        }

        $R \gets (l.\mathrm{core} \cup r.\mathrm{core}) \setminus \{l.\mathrm{lit}, r.\mathrm{lit}\}$\;

        \If{$R = \emptyset$}{
            $\mathrm{close}(root, R)$ \tcp*[r]{global UNSAT}
            \Return\;
        }

        $\mathrm{close}(p, R)$\;
        $cur \gets p$\;
    }
}
\end{algorithm}

\subsection{Asynchronous Core Minimization}\label{subsec:core_min}
A core returned from the SMT engine is not necessarily minimal, and smaller cores enable more powerful pruning. To increase the power of Algorithm \ref{alg:backtracking}, we introduce a dedicated asynchronous \emph{core minimization thread}. When worker $w$ closes node $n$, it submits the pair $(n, C)$, where $C$ is the unrefined UNSAT core, to the core minimization thread's pending queue. The minimizer attempts to refine each enqueued core $C$ by iteratively removing literals $\ell$: it flips $\ell$ to $\neg \ell$, and checks if this preserves UNSAT in under 5000 conflicts (Algorithm~\ref{alg:core_min}). If so, it discards $\ell$, since UNSAT does not depend on $\ell$. This is the standard deletion-based core (MUS/MUC) extraction paradigm \cite{mus}, adapted to bounded SMT checks. If $C$ is reduced, the pair $(n, C)$ is sent to the batch manager for additional search tree pruning via core strengthening (Section \ref{subsec:pruning}).

Note that we must enqueue the pair $(n, C)$ rather than $C$ alone, since cores are minimized relative to the cube (node) under which they were derived. To ensure soundness, we store the associated node $n$ with each core and use $C$ to strengthen the core at $n$ (Algorithm \ref{alg:backtracking}), which leads to more powerful pruning. 
Note that using a separate thread to minimize cores unblocks worker threads from exploring the search tree.

\begin{algorithm}[t]
\caption{Partial Core Minimization}
\label{alg:core_min}

\KwIn{UNSAT core C, SMT input formula $\varphi$}
\KwOut{Reduced core C or a satisfying model}

$\Lambda \gets \mathrm{C}$\tcp*{literals to test}
$\mathrm{mus} \gets \emptyset$\tcp*{min UNSAT core}

\tcp{Invariant: $\varphi \land \mathrm{mus} \land \Lambda$ is UNSAT}

\While{$\Lambda \neq \emptyset$}{
    $\ell \gets \mathrm{last}(\Lambda)$\;
    $\Lambda \gets \Lambda \setminus \{\ell\}$\;
    $\Gamma \gets \mathrm{mus} \cup \Lambda \cup \{\neg \ell\}$ \tcp*{flip $\ell$}

    $\Phi \gets \varphi \cup \{\{\ell\} \mid \ell \in \Gamma\}$\;
    $r \gets \mathrm{check\_sat}(\Phi)$\; 

    \uIf{$r = \mathrm{UNDEF}$}{
        $\mathrm{mus} \gets \mathrm{mus} \cup \{\ell\}$\;
    }
    \uElseIf{$r = \mathrm{SAT}$}{
        $\mathrm{set\_global\_sat}(\mathrm{model(\Phi)})$\;
        \Return\;
    }
    \ElseIf{$r = \mathrm{UNSAT}$}{
        $C \gets \mathrm{UnsatCore}(\Gamma)$\;
        
        \If{$\neg \ell \notin C$}{
            $\Lambda \gets \emptyset$; \quad $\mathrm{mus}' \gets \emptyset$\;

            \ForEach{$c \in C$}{
                \eIf{$c \in \mathrm{mus}$}{
                    $\mathrm{mus}' \gets \mathrm{mus}' \cup \{c\}$\;
                }{
                    $\Lambda \gets \Lambda \cup \{c\}$\;
                }
            }

            $\mathrm{mus} \gets \mathrm{mus}'$\;
        }
    }
}

$\mathrm{C} \gets \mathrm{mus} \cup \Lambda$\;

\end{algorithm}


\subsection{Terminate on Demand}~\label{subsec:terminate_on_demand}
Recall that our node selection policy assigns workers to \emph{active} nodes when no \emph{open} nodes remain, resulting in a portfolio of $m$ workers operating on the same node $n$. If a worker $w_1$ proves $n$ UNSAT, the batch manager closes $n$ and prunes the tree. Consequently, the remaining workers $w_2,\dots,w_m$, as well as any workers exploring pruned regions of the search space, may now be working on stale subproblems. To eliminate such redundant work, similar to \cite{ariparti}, we include a \emph{terminate-on-demand} mechanism based on \emph{node leases}. When worker $w$ is assigned node $n$, it acquires a lease that records the node’s current \emph{cancel epoch}. Each node maintains its own cancel epoch, which is incremented when the node is closed. After pruning, the batch manager terminates all workers with stale leases (i.e. whose recorded cancel epoch no longer matches that of its assigned node), eagerly aborting their current search and reassigning them to new nodes.


\subsection{Online Backbone Detection}~\label{subsec:failed_literals}

We further optimize our architecture with two dedicated \emph{online backbone detection} threads as a lightweight search-space pruning mechanism. To our knowledge, this constitutes the first application of backbone detection to parallel SMT search. The \emph{backbone} of a formula $\varphi$ is the set of literals true in all models of $\varphi$ \cite{janota_bb}. In other words, a literal $\ell$ is in the backbone of $\varphi$ iff $\varphi \models \ell$, i.e. iff $\varphi \land \neg \ell$ is UNSAT. Thus, $\neg \ell$ can prune the partition tree: if $\neg \ell$ matches node $n$'s literal, Algorithm~\ref{alg:backtracking} can close $n$ with singleton core $\{\neg \ell\}$. Typical backbone detection algorithms assume $\varphi$ is SAT and derive candidates from implicants \cite{janota_bb}. In contrast, we extract backbone candidates dynamically based on phase age and evaluate them via two parallel threads operating in complementary modes. In \emph{negative mode}, we negate candidates and attempt to quickly prove them as backbones by deriving a short UNSAT proof. If $\varphi$ is UNSAT, every literal is vacuously in the backbone since $\varphi$ has no models. Even so, proving an individual backbone literal under a small bounded conflict budget is often much easier than proving $\varphi$ UNSAT. In \emph{positive mode}, we assume candidates directly and attempt to extend the assignment to a full model. Theoretically, the positive mode may accelerate search on SAT instances.

Throughout solving, the batch manager stores proven backbone literals and a dynamic ranking of the top 100 candidates from worker threads. Candidates are ranked based on \emph{phase age} (number of assignments since their last phase change) and \emph{cube hits} (number of times worker threads returned them):
\[
\mathrm{rank}(c) = \mathrm{phase\_age}(c) \cdot \log_2\!\bigl(2 + \mathrm{cube\_hits}(c)\bigr)
\]
Candidates are dispatched in batches $\Lambda$ to the backbone threads, who attempt to prove them using a modified version of the chunked backbone algorithm~\cite{janota_bb} (Algorithm \ref{alg:backbones}).\footnote{Algorithm 7 in~\cite{janota_bb}} Given $\Lambda$, the procedure repeatedly selects a chunk $\Gamma \subseteq \Lambda$ of backbone candidates of size at most $K$, and attempts to refute the simultaneous negation of all literals in $\Gamma$ (i.e. it invokes the solver under the assumptions $\omega_N = \{\neg \ell \mid \ell \in \Gamma\}$). If the resulting formula is UNSAT, the UNSAT core $C$ is intersected with the current assumptions $\omega_N$. If this intersection is a singleton $\{\neg \ell\}$, then $\ell$ is a backbone literal. The batch manager collects $\ell$ for search-space pruning and shares it as a unit clause. Otherwise, all literals in $C$ are removed from $\omega_N$, and the refinement loop continues. In our setting, a satisfying assignment under $\omega_N$ signals immediate termination and global SAT; we do not further refine $\Lambda$. Our initial chunk size is 20, and our per-chunk \texttt{check\_sat} conflict budget is 1000.

Algorithm \ref{alg:backbones} falls back to testing individual literals in $\Gamma$ with \emph{failed literal probing} when $\omega_N = \emptyset$ (i.e. when core refinement exhausts the current assumption set without producing a singleton core) or when the maximum-sized chunk ($k = |\Lambda|$) returns UNDEF. A literal $\ell$ is said to be \emph{failed} in a CNF formula $\varphi$ if unit propagation derives a contradiction from $\varphi \land \ell$~\cite{failed_literals}. If $\ell$ is failed, then its complement $\neg \ell$ belongs to the backbone of $\varphi$.
For each $\ell \in \Gamma$, our fallback algorithm checks if $\varphi \land \ell$ is UNSAT in under 10 conflicts; if so, $\neg\ell$ is a backbone literal.

\begin{algorithm}[t]
\caption{Backbone Detection (based on \cite{janota_bb})}
\label{alg:backbones}
\SetKw{Break}{break}
\KwIn{Backbone candidate literals $\Lambda$; SMT input formula $\varphi$; $K \in \mathbb{N}^+$ (chunk size)}

$k \gets \min(K, |\Lambda|)$\tcp*{Initial chunk size}

\While{$\Lambda \neq \emptyset$}{
    $\Gamma \gets$ pick $k$ literals from $\Lambda$\;
    $\omega_N \gets \{\neg{\ell} \mid \ell \in \Gamma\}$\;

    \While{$\mathrm{true}$}{
        $\Phi \gets \varphi \cup \{\{\ell\} \mid \ell \in \omega_N\}$\;
        $r \gets \mathrm{check\_sat}(\Phi)$\;

        \uIf{$r = \mathrm{SAT}$}{
            $\mathrm{set\_global\_sat}(\mathrm{model(\Phi)})$\;
            \Return\;
        }
        \uElseIf{$r = \mathrm{UNSAT}$}{
            $C \leftarrow \mathrm{UnsatCore}(\omega_N)$\;
            \uIf{$C = \emptyset$}{
                $\mathrm{set\_global\_unsat}()$\;
                \Return\;
            }\ElseIf{$C = \{\ell\}$}{
                $\mathrm{collect\_backbone}(\neg\ell)$\;
                $\Lambda \gets \Lambda \setminus \{\neg\ell\}$\;
                $\varphi \gets \varphi \cup \{\neg\ell\}$\;
                $\Gamma \gets \Gamma \setminus \{ \neg\ell \}$\;
            }
            
            \tcp{Remove literals in the core}
            $\omega_N \gets \omega_N \setminus C$\;

            \If{$\omega_N = \emptyset$}{
                \text{test all $\ell \in \Gamma$ with failed literal probing}\;
                $\Lambda \gets \Lambda \setminus \Gamma$\;
                $k \gets \min(K, |\Lambda|)$\tcp*{reset chunk size}
                \Break\tcp*{Done with the chunk}
            }
        } \Else (// $r$ = UNDEF) { 
            \uIf{$k < |\Lambda|$}{
                \tcp{retry with larger chunk}
                $k \gets \min(2k, |\Lambda|)$\;
            } \Else{
                \tcp{Done with the chunk}
                \text{test all $\ell \in \Gamma$ with failed literal probing}\;
                $\Lambda \gets \Lambda \setminus \Gamma$\;
            }
            \Break\;
        }
    }
}

\end{algorithm}

  \section{Evaluation}~\label{sec:evaluation}

We implemented our approach inside \textsc{z3}\cite{z3}.
All experiments were conducted on a dual-socket server with two AMD EPYC 7763 processors (64 cores per socket, 128 cores, 2.45 GHz clock speed), and 512GB RAM, running Ubuntu 24.04.4 LTS. The International Satisfiability Modulo Theories Competition (SMT-COMP) is the standard platform in the SMT community to compare solver performance \cite{smtcomp}. We show our approach outperforms several state-of-the-art parallel solvers on a diverse set of difficult benchmarks from the SMT-COMP 2025 Parallel Track. We then conduct ablations demonstrating that our approach scales with computational resources and benefits from all system components.

\subsection{SMT-COMP 2025 Parallel Track Benchmarks}

We evaluate our approach against the state-of-the-art parallel solvers \textsc{smts} \cite{smts2026}, \textsc{AriParti} \cite{ariparti}, and \textsc{cvc5} in both portfolio (\textsc{smt-d} \cite{smt-d}) and partitioning (\textsc{cvc5-p} \cite{cvc5_partition}) modes on 6 logics from the SMT-COMP 2025 Parallel Track \cite{smtcomp2025}: \textsf{QF\_NIA}, \textsf{QF\_NRA}, \textsf{QF\_LIA}, \textsf{QF\_LRA}, \textsf{QF\_IDL}, and \textsf{QF\_RDL}. The SMT-COMP Parallel Track curates benchmarks designed to be challenging for parallel solvers. We run each solver with 8 threads and a 20-minute wall-clock timeout per benchmark. 

\subsubsection{Experimental Setup} In line with Figure \ref{fig:architecture}, in addition to the 8 partition tree threads that work on the input formula, our \textsc{z3} parallel setup adds a core minimizer thread (Algorithm \ref{alg:core_min}) and 2 backbone detection threads (Algorithm \ref{alg:backbones}) for a total of 11 threads. For the arithmetic benchmarks, we set \texttt{tactic.default\_tactic=smt}, \texttt{smt.auto\_config=false}. For the difference logics, we set \texttt{smt.auto\_config=true}, \texttt{smt.arith\_solver=4}. The sequential backend solver for our \textsc{z3} parallel architecture is simply the current build of \textsc{z3} in single-threaded mode.

We configure each external solver with the parameters specified in the associated paper. Our \textsc{smts} setup uses the iterative tree partitioning algorithm from \cite{smts2026}. \textsc{OpenSMT} is the base solver with $T = 8$ parallel solver instances. We enable partitioning and lemma sharing. The solver timeout $T_S$ starts at 32 seconds and doubles when elapsed time reaches $4 \cdot T_S$. We use the current version of \textsc{OpenSMT} as the sequential \textsc{smts} baseline. We cannot report nonlinear arithmetic results for \textsc{smts} as \textsc{OpenSMT} does not support these logics. 

Our \textsc{AriParti} setup runs the dynamic variable-level partitioning framework from \cite{ariparti} with a maximum of $T = 8$ concurrent worker tasks. \textsc{AriParti} supports three different backend solvers (\textsc{z3}, \textsc{cvc5}, \textsc{OpenSMT}); for comparison, we select \textsc{z3} to best match our own architecture. \textsc{AriParti} uses a pre-compiled \textsc{z3} binary from 2023 (v4.12.1); we thus run this baseline separately from our own current sequential \textsc{z3} build. 

\textsc{smt-d} and \textsc{cvc5-p} both use \textsc{cvc5}\cite{cvc5} as their sequential backend solvers. We build and run the current version of \textsc{cvc5} as the sequential baseline for \textsc{smt-d} and \textsc{cvc5-p}. Our \textsc{smt-d} (\textsc{cvc5} portfolio mode) setup uses the worker diversity strategy from \cite{smt-d} using a portfolio of $T=8$ independent \textsc{cvc5} instances. Clause sharing (which requires the full \textsc{smt-d} gRPC broker) is omitted. Worker diversity follows the guided randomization strategy of \cite{smt-d}: workers are divided into a \emph{standard cluster} (75\%, $\lfloor\frac{3T}{4}\rfloor$ workers) using default randomness, and a \emph{noisy cluster} (25\%, $\big\lfloor\frac{T}{4}\big\rfloor$ workers) using high randomness to explore parts of the search space that standard workers ignore. Standard workers are populated by cycling through logic-specific base option sets (Table II of \cite{smt-d}) in three passes: (1) base options as-is, (2) base options with flipped decision heuristic (\texttt{--decision=justification} if the logic default is \texttt{internal}, else \texttt{--decision=internal}), and (3) base options with \texttt{--decision=internal} and distinct random seeds. Noisy workers use the same logic-specific base options with \texttt{--decision=internal}, \texttt{--random-freq=0.75}, and distinct random seeds. For logics not specified in Table II (nonlinear arithmetic), we fall back to \textsc{cvc5} defaults with \texttt{--decision=internal}.

Our \textsc{cvc5-p} (partition mode) setup uses the graduated portfolio approach from \cite{cvc5_partition}. Given $T=8$ threads, we use a ranked list of partitioning strategies (default: \texttt{decision-scatter}, \texttt{decision-cube}, following the recommended $m=2$ portfolio of \cite{cvc5_partition}) to construct a graduated portfolio: partition batches are allocated greedily in order of increasing size ($2, 4, 8, \ldots$), cycling through the strategy ranking at each size level until we have $T$ partitions. For each batch, \textsc{cvc5} is invoked sequentially with delays of $t_1 = 3\text{s}$ and $t_2 = 0.1\text{s}$ (following \cite{cvc5_partition}) to generate the partitions. Generated partitions across all batches are then solved concurrently using $T$ \textsc{cvc5} solver instances (multijob scheduling).

\subsubsection{Cross-Solver Results} The sequential backend solvers of each parallel architecture vary significantly in performance. For a fair evaluation, we thus examine the performance delta between the sequential backend solver and 8-thread parallel mode for each architecture. Specifically, we examine the delta in number of solved examples and in PAR-2 scores (the sum of the runtimes for all solved instances, plus twice the timeout value multiplied by the number of unsolved instances \cite{par2}). A lower PAR-2 score indicates better performance. Of note, performance profiles naturally vary across SMT-LIB logics because theory solvers employ different reasoning procedures that interact differently with search-space partitioning. In \textsc{Z3}, the \textsf{NIA} and \textsf{NRA} solvers perform much of their theory reasoning during final checks, while the \textsf{RDL} and \textsf{IDL} solvers perform most theory reasoning while propagating assignments to atomic formulas. The \textsf{LIA} solver uses a combination of propagation-based inference and final checks. Consequently, the cost and timing of theory reasoning (and hence the benefit of search-space partitioning) vary by logic. Our evaluation is thus designed to evaluate our framework across these different solving paradigms.

Table~\ref{tab:smt_comp} summarizes our results. In \textsf{QF\_LRA}, we achieve the largest increase in solved instances and the largest PAR-2 reduction by a substantial margin over all competing solvers. In \textsf{QF\_NIA}, we again achieve the largest PAR-2 reduction and tie with \textsc{AriParti} for the largest gain in solved instances. In \textsf{QF\_LIA}, we again achieve the largest PAR-2 reduction while remaining competitive in solved-instance gains, trailing \textsc{AriParti} by only one benchmark and outperforming all other solvers. On \textsf{QF\_IDL}, our solved-instance delta is likewise only one lower than \textsc{smts}, and our PAR-2 reduction is only slightly smaller. Our architecture substantially improves over sequential performance on \textsf{QF\_IDL}; all other architectures besides \textsc{smts} \emph{regress} relative to their sequential baselines on this logic. \textsf{QF\_NRA} is our weakest logic. \textsc{AriParti} is the clear winner on \textsf{QF\_NRA}, but our architecture remains competitive with the remaining solvers. Results on \textsf{QF\_RDL} are inconclusive, as all architectures show negligible changes. Finally, recall that \textsc{AriParti} is specifically designed for arithmetic benchmarks by performing arithmetic-theoretic variable-level partitioning\cite{ariparti}.
Our architecture is highly generalized and not tailored to arithmetic; we nonetheless remain competitive with \textsc{AriParti} on arithmetic benchmarks and also deliver strong results on difference logic (\textsf{QF\_IDL}).

\begin{table*}[t]
\centering
\caption{SMT-COMP 2025 Benchmark Results}
\label{tab:smt_comp}
\scriptsize
\setlength{\tabcolsep}{4pt}

\begin{subtable}[t]{0.49\textwidth}
\centering
\resizebox{\linewidth}{!}{%
\begin{tabular}{ll|ccccc}
\toprule
 &  & \textbf{\textsc{z3} (ours)} & \textsc{smts} & \textsc{AriParti} & \textsc{smt-d} & \textsc{cvc5-p} \\
\midrule

\multirow{6}{*}{\textsf{QF\_LRA (38)}} & Solved (Seq) & 1 & 22 & 6 & 3 & 3 \\
 & PAR-2 (Seq) & 89.37 & 46.10 & 80.70 & 85.80 & 85.80 \\
 & Solved (8T) & 4 & 22 & 7 & 5 & 4 \\
 & PAR-2 (8T) & 83.69 & 45.11 & 77.82 & 82.55 & 84.15 \\
 & $\Delta$ Solved & \textbf{+3} & +0 & +1 & +2 & +1 \\
 & $\Delta$ PAR-2 & \textbf{-5.69} & -0.99 & -2.88 & -3.25 & -1.66 \\
\midrule

\multirow{6}{*}{\textsf{QF\_LIA (44)}} & Solved (Seq) & 2 & 10 & 2 & 1 & 1 \\
 & PAR-2 (Seq) & 102.82 & 86.89 & 101.19 & 103.59 & 103.59 \\
 & Solved (8T) & 8 & 14 & 9 & 3 & 2 \\
 & PAR-2 (8T) & 89.25 & 75.32 & 87.83 & 100.14 & 102.19 \\
 & $\Delta$ Solved & +6 & +4 & \textbf{+7} & +2 & +1 \\
 & $\Delta$ PAR-2 & \textbf{-13.56} & -11.57 & -13.37 & -3.45 & -1.40 \\
\midrule

\multirow{6}{*}{\textsf{QF\_NRA (44)}} & Solved (Seq) & 2 & $\dagger$ & 5 & 1 & 1 \\
 & PAR-2 (Seq) & 100.81 & $\dagger$ & 94.71 & 104.38 & 104.38 \\
 & Solved (8T) & 3 & $\dagger$ & 8 & 2 & 2 \\
 & PAR-2 (8T) & 98.40 & $\dagger$ & 86.98 & 101.71 & 101.26 \\
 & $\Delta$ Solved & +1 & $\dagger$ & \textbf{+3} & +1 & +1 \\
 & $\Delta$ PAR-2 & -2.41 & $\dagger$ & \textbf{-7.73} & -2.68 & -3.13 \\

\bottomrule
\end{tabular}
}
\end{subtable}
\hfill
\begin{subtable}[t]{0.49\textwidth}
\centering
\resizebox{\linewidth}{!}{%
\begin{tabular}{ll|ccccc}
\toprule
 &  & \textbf{\textsc{z3} (ours)} & \textsc{smts} & \textsc{AriParti} & \textsc{smt-d} & \textsc{cvc5-p} \\
\midrule

\multirow{6}{*}{\textsf{QF\_NIA (44)}} & Solved (Seq) & 13 & $\dagger$ & 8 & 4 & 4 \\
 & PAR-2 (Seq) & 76.41 & $\dagger$ & 89.65 & 98.46 & 98.46 \\
 & Solved (8T) & 16 & $\dagger$ & 11 & 4 & 3 \\
 & PAR-2 (8T) & 69.12 & $\dagger$ & 82.58 & 98.02 & 99.58 \\
 & $\Delta$ Solved & \textbf{+3} & $\dagger$ & \textbf{+3} & +0 & -1 \\
 & $\Delta$ PAR-2 & \textbf{-7.29} & $\dagger$ & -7.08 & -0.44 & +1.12 \\
\midrule

\multirow{6}{*}{\textsf{QF\_RDL (24)}} & Solved (Seq) & 1 & 0 & 1 & 0 & 0 \\
 & PAR-2 (Seq) & 55.57 & 57.60 & 55.72 & 57.60 & 57.60 \\
 & Solved (8T) & 1 & 0 & 1 & 0 & 0 \\
 & PAR-2 (8T) & 55.64 & 57.60 & 56.10 & 57.60 & 57.60 \\
 & $\Delta$ Solved & +0 & +0 & +0 & +0 & +0 \\
 & $\Delta$ PAR-2 & +0.07 & +0.00 & +0.38 & +0.00 & +0.00 \\
\midrule

\multirow{6}{*}{\textsf{QF\_IDL (45)}} & Solved (Seq) & 14 & 3 & 5 & 2 & 2 \\
 & PAR-2 (Seq) & 82.21 & 101.80 & 98.98 & 104.97 & 104.97 \\
 & Solved (8T) & 18 & 8 & 4 & 1 & 2 \\
 & PAR-2 (8T) & 75.15 & 92.78 & 100.26 & 106.40 & 105.51 \\
 & $\Delta$ Solved & +4 & \textbf{+5} & -1 & -1 & +0 \\
 & $\Delta$ PAR-2 & -7.06 & \textbf{-9.02} & +1.28 & +1.43 & +0.54 \\

\bottomrule
\end{tabular}
}
\end{subtable}

\vspace{0.5em}

\begin{minipage}{0.85\textwidth}
\footnotesize
\textbf{Legend:}
Solved: number of solved benchmarks (sequential and 8-thread parallel);
PAR-2: penalized average runtime (in thousands);
$\Delta$: difference between 8-thread and sequential runs;
$\dagger$: solver does not support this logic.
\end{minipage}

\end{table*}

Table~\ref{tab:avg-deltas} confirms our system delivers the strongest overall performance across the benchmark suite. We report overall SAT, UNSAT, and total solved-instance deltas between sequential and 8-thread parallel mode for each architecture. Since \textsc{smts} does not support nonlinear arithmetic (2 of the 6 benchmark logics), we additionally report average deltas to enable a fair comparison across all solvers. We also report the average PAR-2 reduction (in thousands) relative to sequential performance. Our architecture achieves both the \textbf{largest average solved-instance gain} and the \textbf{largest average PAR-2 reduction} overall. In particular, we have the \textbf{largest gain in UNSAT examples} (\textsc{smts} wins for SAT examples by a small margin; we tie closely behind with \textsc{AriParti}). The strong gains in both our raw and average UNSAT deltas speak to the effectiveness of our core-guided search-space pruning procedure (Algorithm~\ref{alg:backtracking}).

\begin{table}[t]
\centering
\caption{Aggregate SMT-COMP 2025 Deltas}
\label{tab:avg-deltas}
\resizebox{\columnwidth}{!}{%
\begin{tabular}{l|ccccc}
\toprule
 & \textbf{\textsc{z3} (ours)} & \textsc{smts} & \textsc{AriParti} & \textsc{smt-d} & \textsc{cvc5-p} \\
\midrule
$\Delta$ SAT & +12 & +9\textbf{*} & +12 & +3 & +3 \\
$\Delta$ UNSAT & +5 & +0\textbf{*} & +1 & +1 & -1 \\
$\Delta$ All & +17 & +9\textbf{*} & +13 & +4 & +2 \\
\midrule
Avg. $\Delta$ SAT & +2.00 & \textbf{+2.25} & +2.00 & +0.50 & +0.50 \\
Avg. $\Delta$ UNSAT & \textbf{+0.83} & +0.00 & +0.17 & +0.17 & -0.17 \\
Avg. $\Delta$ All & \textbf{+2.83} & +2.25 & +2.17 & +0.67 & +0.33 \\
\midrule
Avg. $\Delta$ PAR-2 & \textbf{-5.99} & -5.39 & -4.90 & -1.40 & -0.76 \\
\bottomrule
\end{tabular}
}
\par\vspace{0.5em}
\begin{minipage}{\columnwidth}
\footnotesize
\textbf{*}\textsc{smts} does not support NIA, NRA; the rest handle all 6 logics. PAR-2 score is in thousands.
\end{minipage}
\end{table}

\subsection{Ablations}
The SMT-COMP Parallel benchmark set is relatively small and designed for cross-solver comparison, making it less suitable for isolating the effects of individual algorithmic components. Larger, balanced, and more diverse benchmark suites yield more robust estimates of each component’s contribution. We thus run ablations of our system on randomly selected subsets from the 2024 and 2025 SMT-LIB \textsf{QF\_LIA} and \textsf{QF\_NIA} nonincremental benchmark suites \cite{smtlib}. Each experiment uses a 10-minute (600-second) per-benchmark timeout. We demonstrate (1) how our system scales with computational resources and (2) the importance of the core-guided pruning mechanism, core minimizer thread, and backbone threads.

\subsubsection{Scalability} 
\begin{figure}[!t]
    \centering

    \begin{minipage}{0.35\textwidth}
        \centering
        \includegraphics[width=\linewidth]{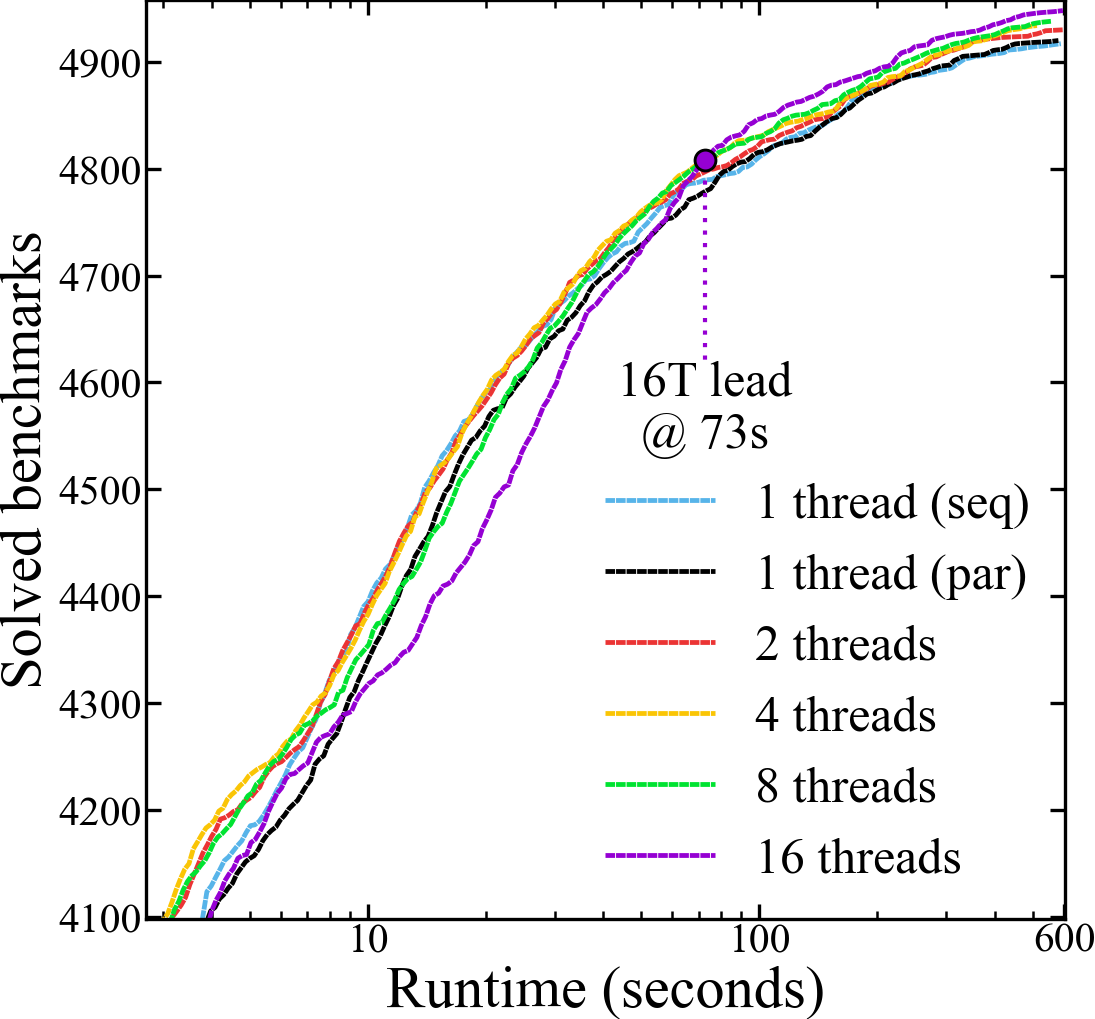}

        {\small (a) QF\_LIA}
    \end{minipage}
    
    \vspace{0.5em}
    
    \begin{minipage}{0.35\textwidth}
        \centering
        \includegraphics[width=\linewidth]{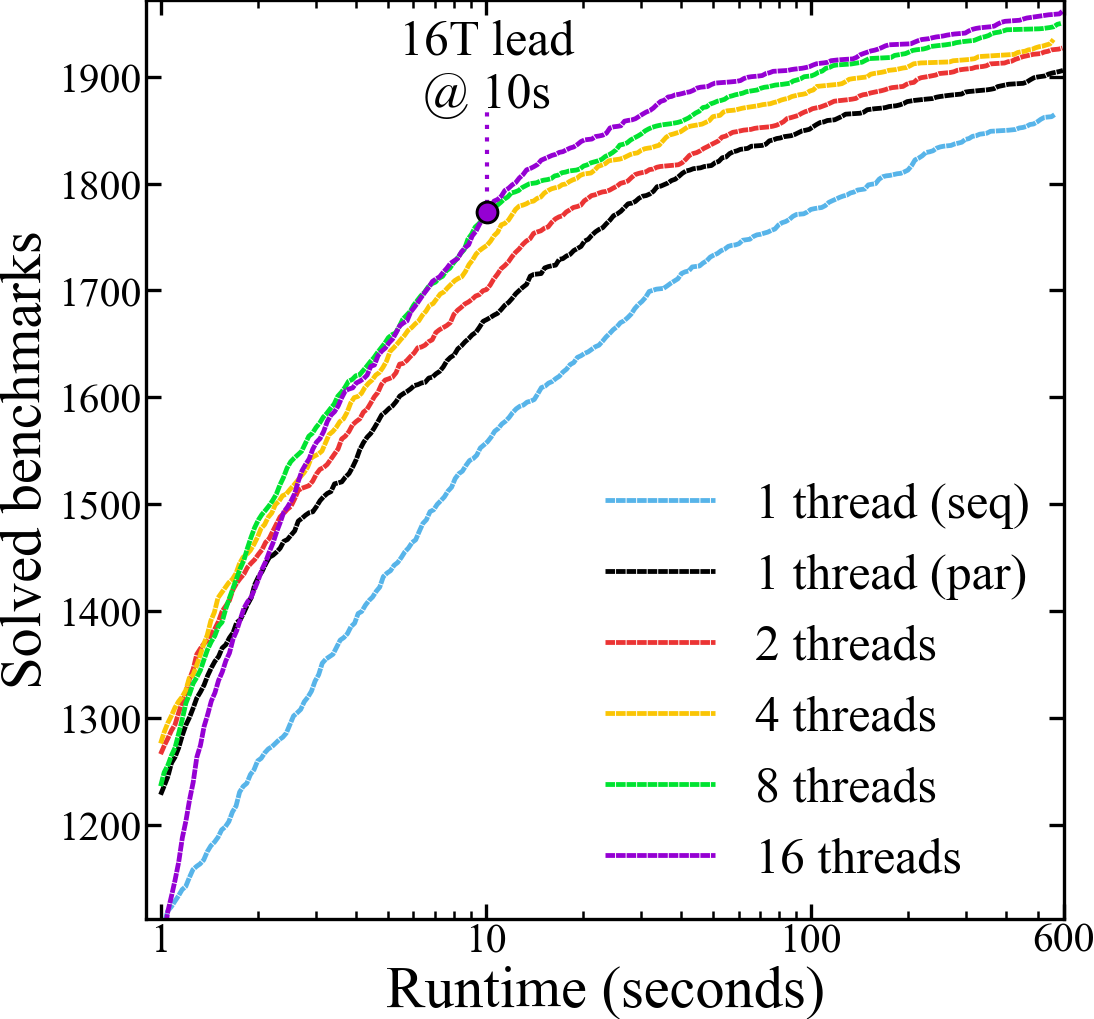}

        {\small (b) QF\_NIA}
    \end{minipage}

    \caption{Scaling of Parallel Z3 Using 1 to 16 Solvers}
    \label{fig:scaling}
\end{figure}

To show how our system scales with available resources, we run it with 1 (vanilla sequential \textsc{z3}), 1 (using our parallel framework), 2, 4, 8, and 16 worker threads on two randomly selected subsets from the complete 2024 SMT-LIB nonincremental benchmark suite \cite{smtlib2025} (consisting of 5000 and 2313 benchmarks, respectively, for \textsf{QF\_LIA} and \textsf{QF\_NIA}). Figure \ref{fig:scaling} summarizes our results via cactus plots. We use a logarithmic x-axis for runtime in seconds (1 to 600), and a y-axis representing the cumulative number of benchmarks each configuration solved under the 600-second timeout. On \textsf{QF\_LIA}, easy benchmarks do not benefit from parallelization, likely due to concurrency overhead on otherwise simple problems. Parallelism becomes beneficial on harder instances, with 16-threads taking the lead after 73 seconds. The benefit is even more pronounced for \textsf{QF\_NIA}: sequential execution performs substantially worse than every parallel configuration, and performance scales smoothly with thread count, with 16-threads taking the lead after only 10 seconds. The increased benefit of our framework on \textsf{QF\_NIA} is consistent with its greater practical and theoretical difficulty: \textsc{z3} performance on \textsf{QF\_NIA} is unstable in practice, and while \textsf{QF\_LIA} satisfiability is NP-complete \cite{lia_npcomplete}, \textsf{QF\_NIA} satisfiability is undecidable \cite{nia_undecidable}. Search-space partitioning is thus beneficial when concurrency overhead, which increases with the thread count, is amortized by the reduced complexity of the resulting subproblems.

\subsubsection{Solver Ablations} The remaining ablations use a carefully balanced subset of the 2025 SMT-LIB nonincremental \textsf{QF\_LIA} benchmark suite \cite{smtlib2024}, consisting of 800 benchmarks (400 SAT / 400 UNSAT) up to 1.2MB in size each. Trivial instances solved by sequential \textsc{z3} in under 10 seconds are excluded. To account for inherent variability in runtimes, we repeat each ablation five times, and report the mean differences in both PAR-2 score (in thousands) and number of solved instances relative to the full configuration. We further visualize the pooled results across \emph{all} five repetitions of each ablation using scatter plots, where {\color{green}$\times$} denotes a SAT benchmark solved by either configuration, {\color{red}$\square$} denotes an UNSAT benchmark, and {\color{blue}$\ast$} denotes a timed-out benchmark. The axes use a logarithmic scale for per-benchmark runtimes in the range $[0,600]$ seconds; the diagonal denotes equal runtimes for both configurations.

\begin{figure}[!t]
    \centering
    \includegraphics[width=0.49\textwidth]{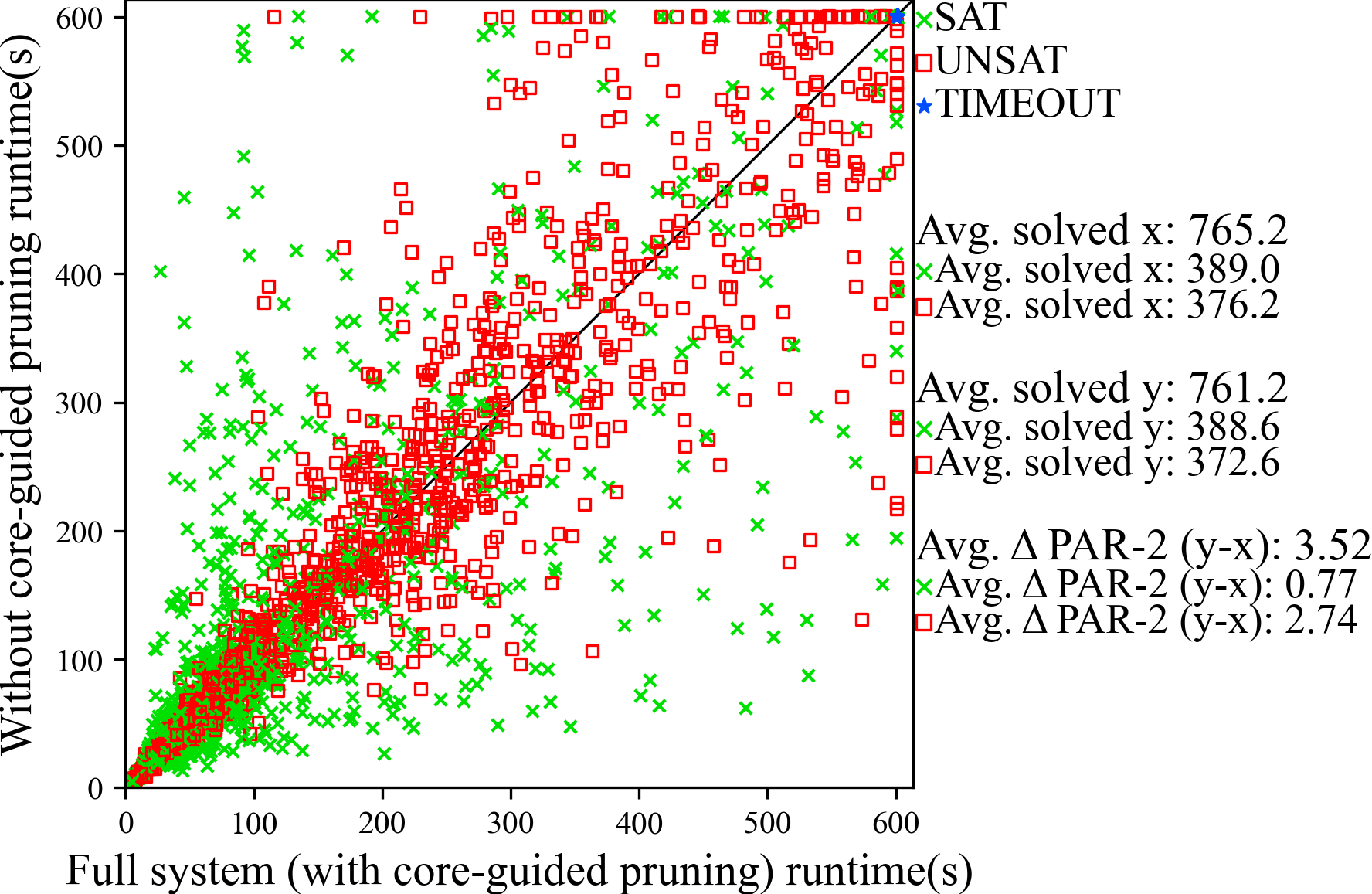}
    \caption{Core-Guided Pruning Ablation (5 Runs, Pooled)}
    \label{fig:ablate_backtracking}
\end{figure}

\emph{Core-Guided Pruning.} To ablate core-guided pruning (Algorithm \ref{alg:backtracking}), when closing node $n$, we ignore $n$'s derived core and instead use $n$'s full cube (path from root to $n$) as its core. This prunes only $n$ and its subtree, similar to \textsc{smts} and \textsc{AriParti}. Removing core-guided pruning also renders the core minimizer thread obsolete, so we disable it. This ablation addresses the only fully synchronous component of our architecture: core-guided pruning modifies shared tree state and must therefore execute under a global lock, temporarily blocking the batch manager. 
Our results show that the benefit of pruning outweighs this synchronization overhead (Table \ref{tab:backtracking-ablation}), with the results pooled across all five runs visualized in Figure \ref{fig:ablate_backtracking}. Enabling core-guided pruning solves an average of four additional benchmarks and decreases the mean PAR-2 score by 3.52 points. UNSAT benchmarks account for an average of 3.60 (90\%) of the additional solves and 2.74 points (78\%) of the PAR-2 reduction, supporting the hypothesis that core-guided pruning particularly benefits UNSAT problems by dynamically shrinking the search space.

\begin{table}[t]
\centering
\caption{Core-Guided Pruning Ablation Aggregates}
\label{tab:backtracking-ablation}
\resizebox{\columnwidth}{!}{%
\begin{tabular}{l|cc}
\toprule
\textbf{Metric} & \begin{tabular}[t]{@{}c@{}}\textbf{Full system} \\ \textbf{(with core-guided pruning)}\end{tabular} & \begin{tabular}[t]{@{}c@{}}\textbf{No core-} \\ \textbf{guided pruning}\end{tabular} \\
\midrule
Avg. SAT & 389.0 & 388.6 \\
Avg. UNSAT & 376.2 & 372.6 \\
Avg. total solved & 765.2 & 761.2 \\
Avg. PAR-2 & 123.83 & 127.35 \\
\hdashline
Avg. $\Delta$ SAT & -0.00 & -0.40 \\
Avg. $\Delta$ UNSAT & -0.00 & -3.60 \\
Avg. $\Delta$ solved & -0.00 & -4.00 \\
Avg. $\Delta$ PAR-2 & +0.00 & +3.52 \\
\bottomrule
\end{tabular}
}
\par\vspace{0.5em}
\begin{minipage}{\columnwidth}
\footnotesize
PAR-2 score is in thousands. Aggregates are over 5 runs for each mode.
\end{minipage}
\end{table}

\begin{figure}[!t]
    \centering
    \includegraphics[width=0.49\textwidth]{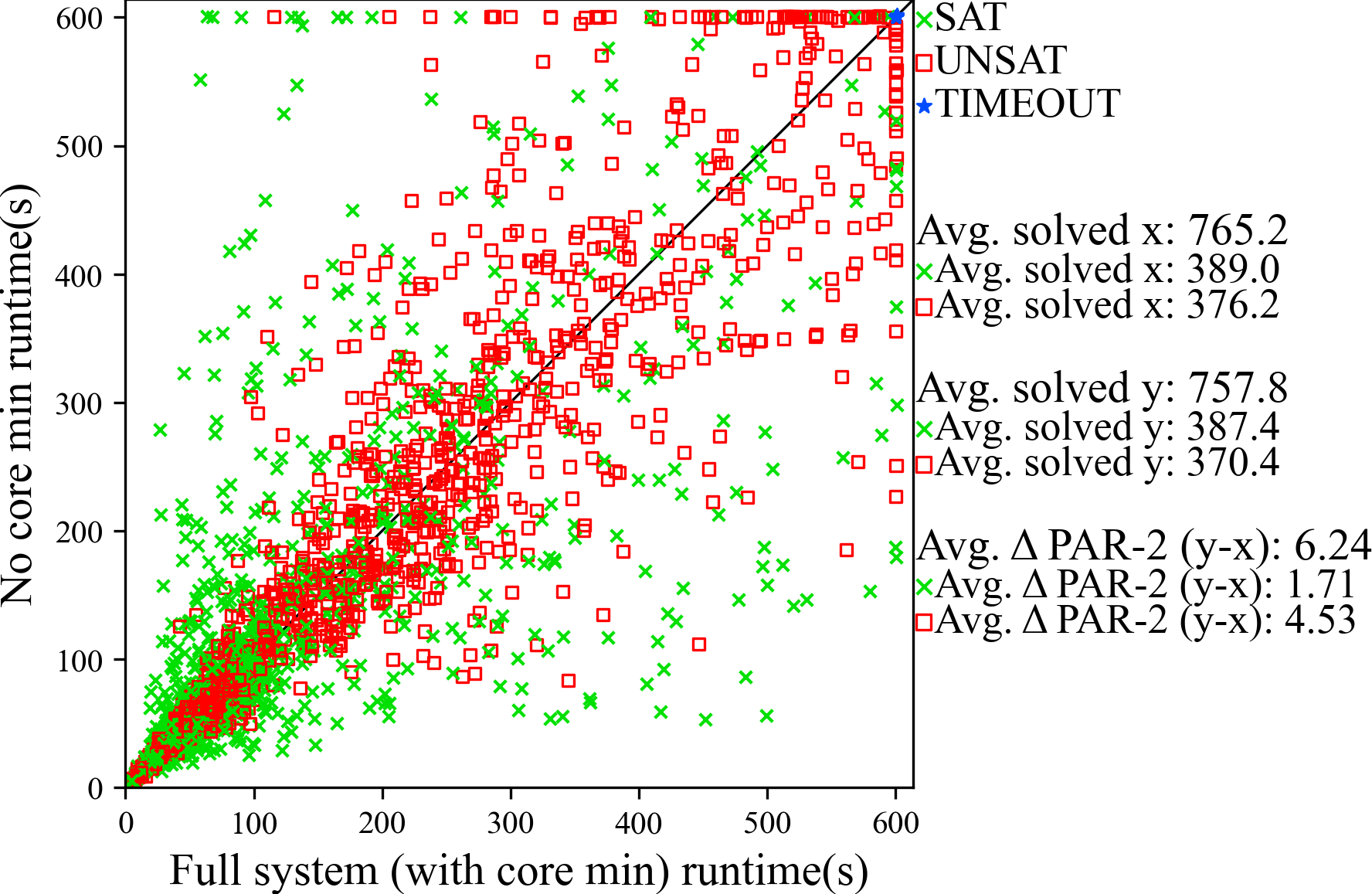}
    \caption{Core Minimization Ablation (5 Runs, Pooled)}
    \label{fig:ablate_core_min}
\end{figure}

\emph{Core Minimizer Thread.} We next ablate the core minimizer thread. Across five runs, enabling core minimization solves an average of 7.4 additional benchmarks and decreases the mean PAR-2 score by 6.24 points (Table \ref{tab:core-min-ablation}). Of these additional solves, an average of 5.8 (78\%) are UNSAT, supporting the hypothesis that core minimization particularly benefits UNSAT problems by enabling more aggressive search-space pruning. Figure \ref{fig:ablate_core_min} visualizes the results pooled across all five runs. Despite per-instance variability across runs, reflected by the dispersion around the diagonal, core minimization reduces overall PAR-2 by 5\%, with reductions of 4\% for SAT and 5\% for UNSAT benchmarks.

\emph{Core Minimizer Thread.} We next ablate the core minimizer thread. The aggregate results are reported in Table \ref{tab:core-min-ablation}, with the results pooled across all five runs visualized in Figure \ref{fig:ablate_core_min}. Enabling core minimization solves an average of 7.4 additional benchmarks and decreases the mean PAR-2 score by 6.24 points. UNSAT benchmarks account for an average of 5.8 (78\%) of the additional solves and 4.53 points (73\%) of the PAR-2 reduction, supporting the hypothesis that core minimization particularly benefits UNSAT problems by strengthening the cores used by core-guided pruning, thereby enabling more aggressive search-space pruning.

\begin{table}[t]
\centering
\caption{Core Min Ablation Aggregates}
\label{tab:core-min-ablation}
\resizebox{0.8\columnwidth}{!}{%
\begin{tabular}{l|cc}
\toprule
\textbf{Metric} & \begin{tabular}[t]{@{}c@{}}\textbf{Full system} \\ \textbf{(with core min)}\end{tabular} & \begin{tabular}[t]{@{}c@{}}\textbf{No core} \\ \textbf{minimization}\end{tabular} \\
\midrule
Avg. SAT & 389.0 & 387.4 \\
Avg. UNSAT & 376.2 & 370.4 \\
Avg. total solved & 765.2 & 757.8 \\
Avg. PAR-2 & 123.83 & 130.07 \\
\hdashline
Avg. $\Delta$ SAT & -0.00 & -1.60 \\
Avg. $\Delta$ UNSAT & -0.00 & -5.80 \\
Avg. $\Delta$ solved & -0.00 & -7.40 \\
Avg. $\Delta$ PAR-2 & +0.00 & +6.24 \\
\bottomrule
\end{tabular}
}
\par\vspace{0.5em}
\begin{minipage}{\columnwidth}
\footnotesize
PAR-2 score is in thousands. Aggregates are over 5 runs for each mode.
\end{minipage}
\end{table}

\begin{figure}[!t]
    \centering

    \begin{minipage}{0.49\textwidth}
        \centering
        \includegraphics[width=\linewidth]{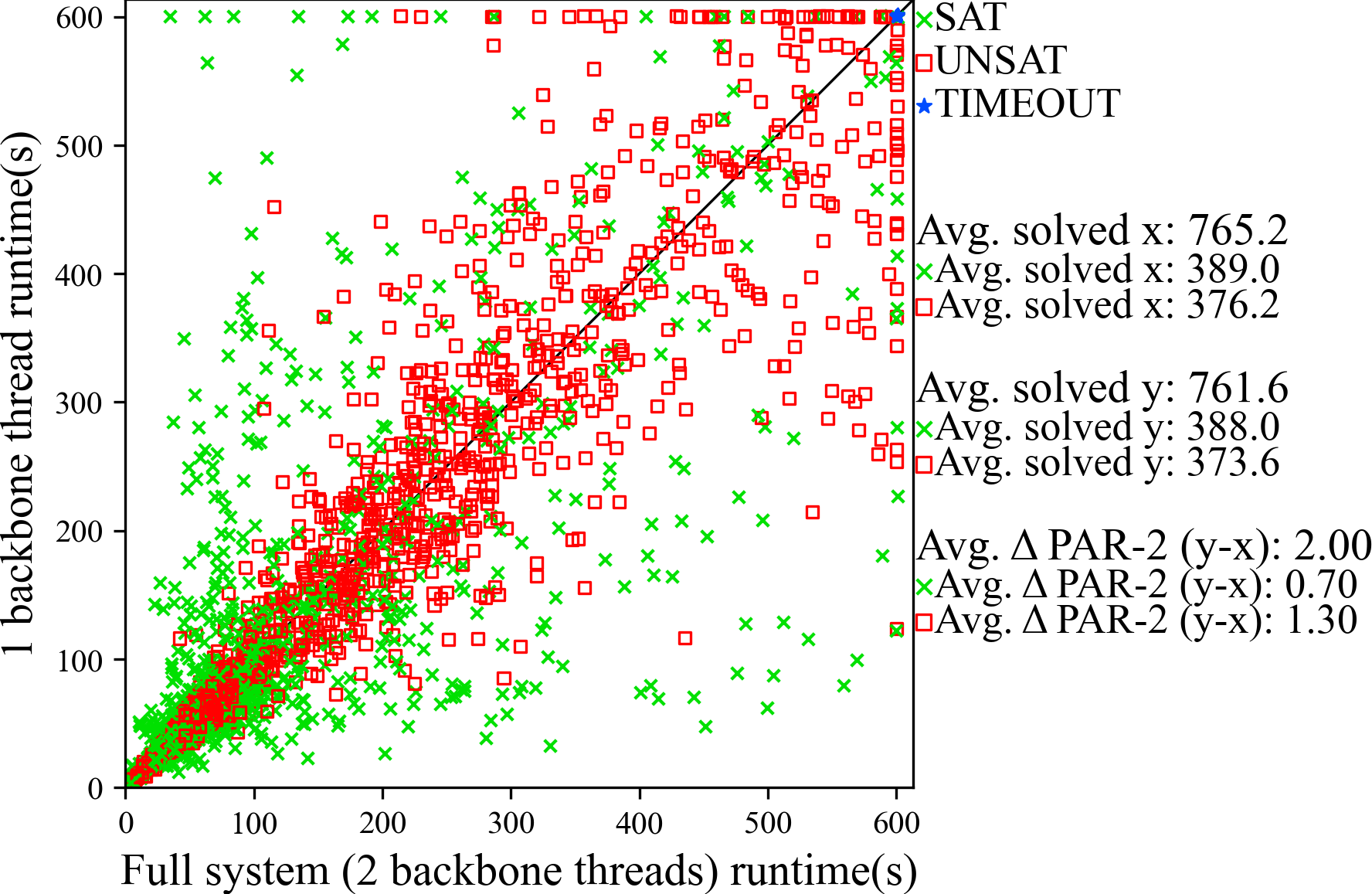}

        {\small (a) 2 vs 1 Backbone Threads}
    \end{minipage}

    \vspace{0.5em}

    \begin{minipage}{0.49\textwidth}
        \centering
        \includegraphics[width=\linewidth]{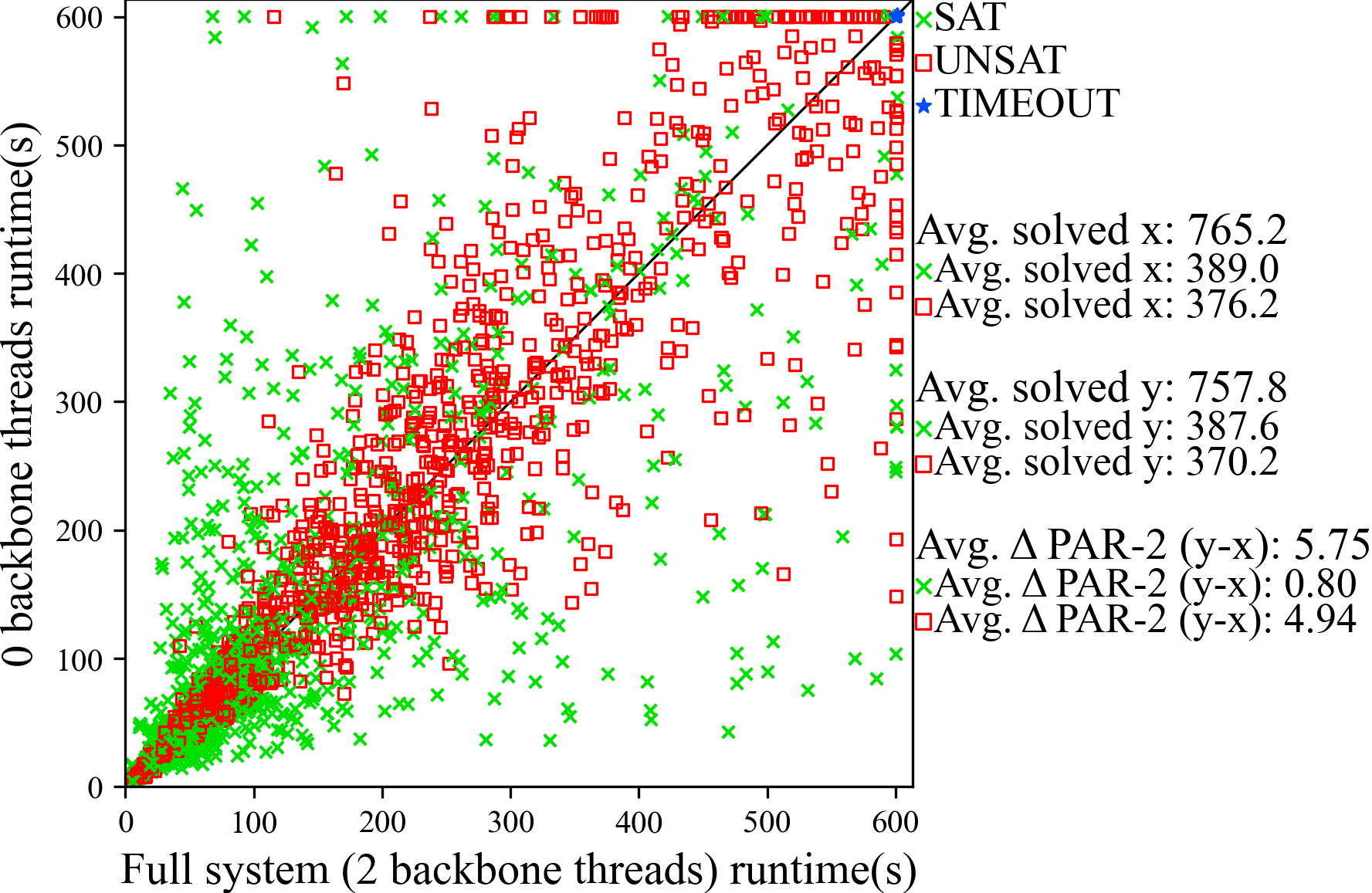}

        {\small (b) 2 vs 0 Backbone Threads}
    \end{minipage}

    \begin{minipage}{0.49\textwidth}
        \centering
        \includegraphics[width=\linewidth]{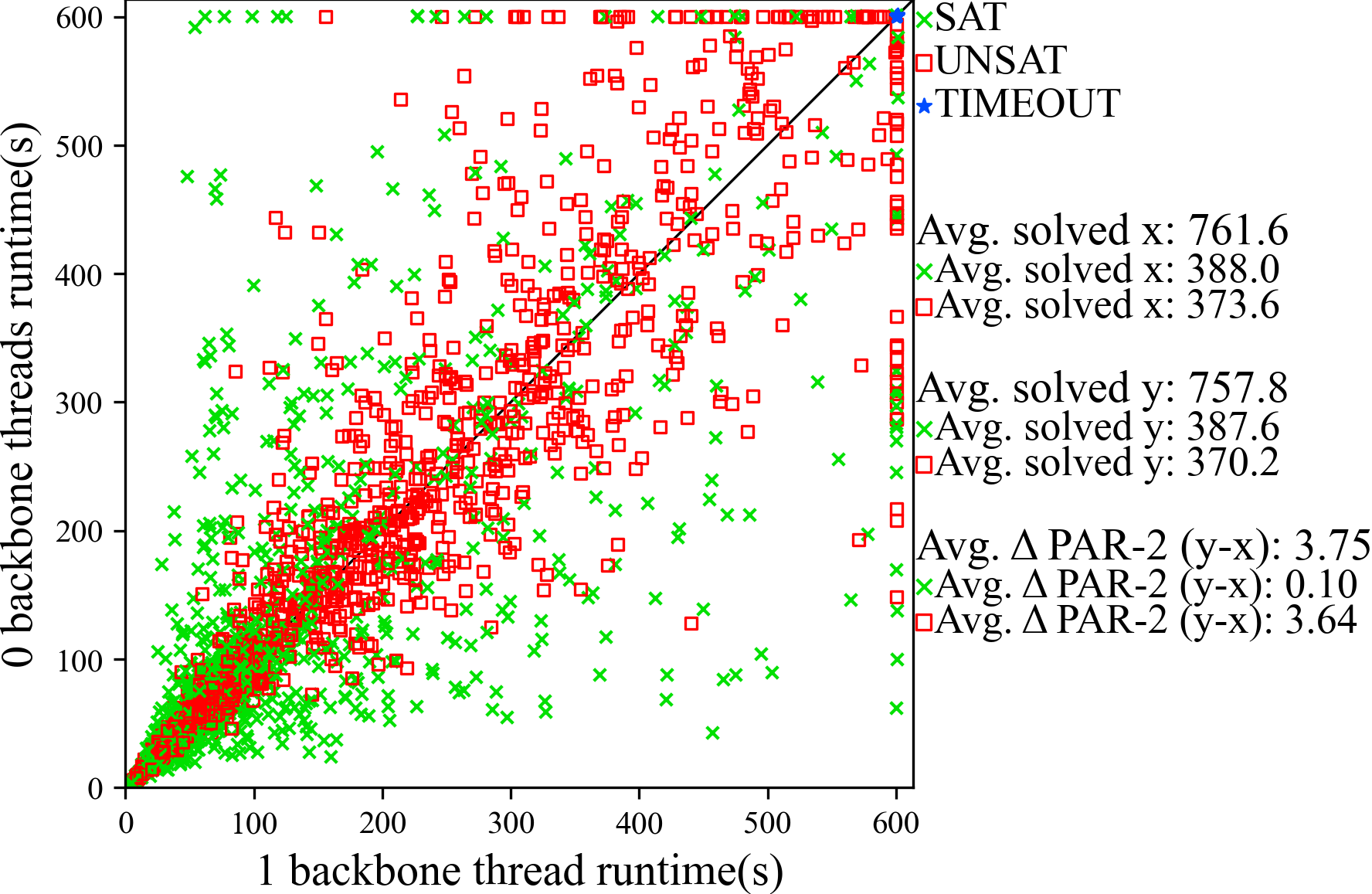}

        {\small (c) 1 vs 0 Backbone Threads}
    \end{minipage}

    \caption{Backbone Thread Ablations (5 Runs, Pooled)}
    \label{fig:ablate_backbones}
\end{figure}

\emph{Backbone Threads.} 
Recall that backbone detection proves that a literal $\ell$ belongs to the backbone by showing that $\varphi \land \neg \ell$ is UNSAT; we call this \emph{negative mode} \cite{janota_bb}. Our bidirectional online backbone detection architecture complements this with a \emph{positive mode}, which assumes candidate literals directly and attempts to extend the resulting assignment into a full model. We evaluate the full system with 2 backbone threads (negative and positive modes) against configurations with 1 thread (negative mode only) and 0 backbone threads. The aggregate results are reported in Table \ref{tab:backbone-ablation}, with the results pooled across all five runs visualized in Figure \ref{fig:ablate_backbones}. 

Performance improves with each added backbone thread. Moving from 0 to 1 backbone thread solves an average of 3.8 additional benchmarks and decreases mean PAR-2 by 3.75 points; UNSAT benchmarks account for 3.4 (89\%) of these additional solves and 3.64 points (97\%) of the PAR-2 reduction. Moving from 1 to 2 backbone threads solves another 3.6 benchmarks and decreases mean PAR-2 by a further 2.00 points; UNSAT benchmarks account for 2.6 (72\%) of these additional solves and 1.30 points (65\%) of the additional PAR-2 reduction. Although both transitions favor UNSAT, positive mode yields larger incremental SAT gains than negative mode: +1.0 vs +0.4 average solved instances, and -0.70 vs -0.10 PAR-2 points. This is consistent with our hypothesis that positive mode facilitates SAT search. Overall, the full setup solves 7.4 more benchmarks and decreases mean PAR-2 by 5.75 points relative to 0 backbone threads. UNSAT benchmarks account for 6.0 (81\%) of these total additional solves and 4.94 points (86\%) of the total PAR-2 reduction. This strong UNSAT skew is consistent with proven backbone literals shrinking the search space and accelerating exhaustive UNSAT proofs, while the incremental SAT gains from positive mode demonstrate the complementary benefit of the full dual-mode architecture.

\begin{table}[t]
\centering
\caption{Backbone Ablation Aggregates}
\label{tab:backbone-ablation}
\resizebox{\columnwidth}{!}{%
\begin{tabular}{l|ccc}
\toprule
\textbf{Metric} & \begin{tabular}[t]{@{}c@{}}\textbf{Full system} \\ \textbf{(2 backbone threads)}\end{tabular} & \begin{tabular}[t]{@{}c@{}}\textbf{1 backbone} \\ \textbf{thread}\end{tabular} & \begin{tabular}[t]{@{}c@{}}\textbf{0 backbone} \\ \textbf{threads}\end{tabular} \\
\midrule
Avg. SAT & 389.0 & 388.0 & 387.6 \\
Avg. UNSAT & 376.2 & 373.6 & 370.2 \\
Avg. total solved & 765.2 & 761.6 & 757.8 \\
Avg. PAR-2 & 123.83 & 125.83 & 129.58 \\
\hdashline
Avg. $\Delta$ SAT & -0.00 & -1.00 & -1.40 \\
Avg. $\Delta$ UNSAT & -0.00 & -2.60 & -6.00 \\
Avg. $\Delta$ solved & -0.00 & -3.60 & -7.40 \\
Avg. $\Delta$ PAR-2 & +0.00 & +2.00 & +5.75 \\
\bottomrule
\end{tabular}
}
\par\vspace{0.5em}
\begin{minipage}{\columnwidth}
\footnotesize
PAR-2 score is in thousands. Aggregates are over 5 runs for each mode.
\end{minipage}
\end{table}

  \section{Conclusion}~\label{sec:conclusion}

We introduced a novel framework for parallel SMT solving that \emph{learns from active search} by integrating dynamic VSIDS-based partitioning, core-guided search-space pruning, and online backbone detection into a feedback-driven architecture that uses the evolving search state to steer solving. Our system achieves the strongest overall performance across six challenging SMT-COMP 2025 benchmark suites against four state-of-the-art parallel SMT frameworks, and scales effectively with computational resources.
Future work will investigate additional mechanisms for learning from active search, such as online parameter tuning, improved heuristics for split atom and backbone candidate selection, and tailoring proof strategies to individual subproblems. Our approach naturally extends to quantifiers within \textsc{z3}; we plan to conduct further evaluations on quantified benchmarks. Finally, we intend to incorporate non-unit clause sharing, particularly subtree-guarded approaches that exploit search-space locality.
  \section{Acknowledgements}
We thank Diego Oliver Fernandez Pons for suggesting the idea of using backbone candidates. We also thank the anonymous reviewers for ample constructive feedback.
  
  \bibliographystyle{IEEEtran}
  \bibliography{references}
  
\end{document}